\documentclass[oribibl,envcountsame]{llncs}
\usepackage[T1]{fontenc}
\usepackage{ae}
\usepackage[english]{babel}
\usepackage[psamsfonts]{amsfonts}
\usepackage{stmaryrd,amsmath,amssymb,amsbsy}
\usepackage{graphicx}
\usepackage{galois}
\usepackage{color}
\usepackage{index}
\usepackage{url}

\def\astree{{\sc Astr\'ee}}

\def\mc#1{\mathcal{#1}}
\def\mB#1{\mathbb{#1}}

\def\mi#1{\mathit{#1}}

\def\P{\mc{P}}
\def\D{\mc{D}}

\def\V{\mc{V}}
\def\L{\mc{L}}

\def\S{\mc{C}}

\def\Z{\mB{Z}}
\def\Q{\mB{Q}}
\def\R{\mB{R}}
\def\I{\mB{I}}
\def\X{\mc{X}}

\def\s{\sharp}

\def\widen{\mathbin{\triangledown}}

\def\expleq{\preceq}

\def\trunc{\mi{truncate}}
\def\subst{\mi{subst}}
\def\strat{\mi{strat}}
\def\occ{\mi{occ}}
\def\expr{\mi{expr}}

\def\lb#1{\llbracket\,#1\,\rrbracket}
\def\lc#1{{\{\hskip-0.25em|}\,#1\,{|\hskip-0.25em\}}}
\def\lp#1{\llparenthesis\,#1\,\rrparenthesis}

\def\deq{\;\stackrel{\mbox{{\rm\tiny def}}}{=}\;}
\def\diff{\;\;{\stackrel{\mbox{{\rm\tiny def}}}{\iff}}\;\;}

\def\Int{\mc{I}}

\newtheorem{defn}{Definition}
\newtheorem{thm}{Theorem}

\def\doframeit#1{\hbox{\vbox{%
  \hrule height \fboxrule
    \hbox{%
      \vrule width \fboxrule\kern\fboxsep%
      \vbox{\kern\fboxsep #1 \kern\fboxsep}%
      \kern\fboxsep\vrule width \fboxrule}%
    \hrule height \fboxrule}}}

\def\myfig{\begin{figure}[t]\begin{center}}
\def\endmyfig{\end{center}\end{figure}}

\renewcommand{\paragraph}[1]{\vspace*{0.1cm}\noindent{\em #1}}

\begin{document}


\title{Symbolic Methods to Enhance the Precision of Numerical Abstract 
Domains\thanks{This work was partially supported by the \astree\ RNTL 
project and the APRON project from the ACI ``S\'ecurit\'e \& Informatique.''}}

\author{Antoine Min\'e}

\institute{\'Ecole Normale Sup\'erieure, Paris, France,\\
           \email{mine@di.ens.fr},\\
           \email{http://www.di.ens.fr/$\sim$mine}}

\maketitle


\begin{abstract}
We present lightweight and generic symbolic methods to
improve the precision of numerical static analyses based on 
Abstract Interpretation.
The main idea is to simplify numerical expressions before they are fed to
abstract transfer functions.
An important novelty is that these simplifications are performed on-the-fly,
using information gathered dynamically by the analyzer.

A first method, called ``linearization,'' allows abstracting arbitrary
expressions into affine forms with interval coefficients while simplifying them.
A second method, called ``symbolic constant propagation,'' enhances the
simplification feature of the linearization by propagating assigned
expressions in a symbolic way.
Combined together, these methods increase the relationality level of
numerical abstract domains and make them more robust against
program transformations.
We show how they can be integrated within the classical
interval, octagon and polyhedron domains.

These methods have been incorporated within the \astree\ static analyzer that
checks for the absence of run-time errors in embedded critical avionics 
software.
We present an experimental proof of their usefulness.
\end{abstract}


\section{Introduction}

\begin{myfig}
\begin{tabular}{l}
$X\leftarrow [-10,20]$;\\
$Y\leftarrow X$;\\
if ($Y\leq 0$) \{ $Y\leftarrow -X$; \}\\
{\em // here, $Y\in[0,20]$}\\
\end{tabular}
\caption{Absolute value computation example.}
\label{absolute}
\end{myfig}

\begin{myfig}
\begin{tabular}{l}
$X\leftarrow [0,1]$; \\
$Y\leftarrow [0,0.1]$;\\
$Z\leftarrow [0,0.2]$;\\
$T\leftarrow (X\times Y) - (X\times Z) + Z$;\\
{\em // here, $T\in[0,0.2]$}\\
\end{tabular}
\caption{Linear interpolation computation example.}
\label{interpol}
\end{myfig}

\vspace*{-0.2cm}

Ensuring the correctness of software is a difficult but important task,
especially in embedded critical applications such as planes or rockets.
There is currently a great need for static analyzers able to provide
invariants automatically and directly on the source code.
As the strongest invariants are not computable in general,
such tools need to perform sound approximations at the expense of completeness.
In this article, we will only consider the properties of numerical variables
and work in the Abstract Interpretation framework.
A static analyzer is thus parameterized by a {\em numerical abstract domain}, 
that is, a set of computer-representable numerical properties together with 
algorithms to compute the semantics of program instructions.

There already exist quit a few numerical abstract domains.
Well-known examples include the interval domain \cite{interv} that discovers 
variable bounds, and the polyhedron domain \cite{poly} for affine inequalities.
Each domain achieves some cost versus precision balance.
In particular, non-relational domains---{\em e.g.}, the interval domain---are
much faster but also much less precise than relational domains---able to 
discover variable relationships.
Although the interval information seem sufficient---it allows expressing most 
correctness requirements, such as the absence of arithmetic overflows or 
out-of-bound array accesses---relational invariants are often necessary 
during the course of the analysis to find tight bounds.
Consider, for instance, the program of Fig.~\ref{absolute} that computes the
absolute value of $X$.
We expect the analyzer to infer that, at the end of the program, $Y\in[0,20]$.
The interval domain will find the coarser result $Y\in[-20,20]$ 
because it cannot exploit the information $Y=X$ during the test $Y\leq 0$.
The polyhedron domain is precise enough to infer
the tightest bounds, but results in a loss of efficiency.
In our second example, Fig.~\ref{interpol}, $T$ is linearly interpolated
between $Y$ and $Z$, thus, we have $T\in[0,0.2]$.
Using plain interval arithmetics, one finds the coarser result 
$T\in[-0.2,0.3]$.
As the assignment in $T$ is not affine, the polyhedron
domain cannot perform any better.

In this paper, we present symbolic enhancement techniques that can be applied
to abstract domains to solve these problems and increase their robustness
against program transformations.
In Fig.~\ref{absolute}, our {\it symbolic constant propagation\/} is able 
to propagate the information $Y=X$ and discover tight bounds using
only the interval domain.
In Fig.~\ref{interpol}, our {\em linearization\/} technique allows us
to prove that $T\in[0,0.3]$ using the interval domain (this result 
is not optimal, but still much better than $T\in[-0.2,0.3]$).
The techniques are generic and can be applied to other domains, such as the 
polyhedron domain.
However, the improvement varies greatly from one example to
another and enhanced domains do not enjoy best abstraction functions.
Thus, our techniques depend upon {\em strategies}, some of which are proposed
in the article.

\paragraph{Related Work.}
Our linearization can be related to {\em affine arithmetics\/},
a technique introduced by Vin\'icius et al. in \cite{aa} to refine interval 
arithmetics by taking into account existing correlations between computed 
quantities.
Both use a symbolic form with linear properties to allow basic algebraic
simplifications.
The main difference is that we relate directly program variables while 
affine arithmetics introduces synthetic variables.
This allows us to treat control flow joins and loops, and to interact with
relational domains, which is not possible with affine arithmetics.
Our linearization was first introduced in \cite{mine:float} to abstract
floating-point arithmetics. 
It is presented here with some improvements---including the introduction
of several strategies.

Our symbolic constant propagation technique is similar to the classical
constraint propagation proposed by Kildall in \cite{kildall} to perform
optimization. 
However, scalar constants are replaced with expression trees, and our goal
is not to improve the efficiency but the precision of the abstract
execution.
It is also related to the work of Colby: he introduces, in \cite{colby},
a language of transfer relations to propagate, combine and
simplify, in a fully symbolic way, sequences of transfer functions.
We are more modest as we do not handle disjunctions symbolically and
do not try to infer symbolic loop invariants.
Instead, we rely on the underlying numerical abstract domain to perform most
of the semantical job.
A major difference is that, while Colby's framework statically transforms the 
abstract equation system to be solved by the analyzer, our framework performs 
this transformation on-the-fly and benefits from the information dynamically 
inferred by the analyzer.

\paragraph{Overview of the Paper.}
The paper is organised as follows.
In Sect.~\ref{frameworksect}, we introduce a language---much simplified for 
the sake of illustration---and recall how to perform a numerical static 
analysis parameterized by an abstract domain.
Sect.~\ref{symbolicsect} then explains how symbolic expression manipulations 
can be soundly incorporated within the analysis.
Two symbolic methods are then introduced: expression linearization, 
in Sect.~\ref{linearizationsect}, and symbolic constant propagation, in 
Sect.~\ref{constantsect}.
Sect.~\ref{astreesect} discusses our practical implementation within the 
\astree\ static analyzer and presents some experimental results.
We conclude in Sect.~\ref{conclusionsect}.

\section{Framework}
\label{frameworksect}

\vspace*{-0.2cm}

In this section, we briefly recall the classical design of a static analyzer
using the Abstract Interpretation framework by Cousot and Cousot
\cite{ai,ai3}.
This design is specialised towards the automatic computation of 
{\em numerical\/} invariants, and thus, is parameterized by a numerical 
abstract domain.

\subsection{Syntax of the Language}

\begin{myfig}
\begin{tabular}{cclll}
{\it expr} &::=&  
   $X$ &\qquad\qquad& $X\in\V$ \\
&|& $[a,b]$ && $a\in\I\cup\{-\infty\},\,b\in\I\cup\{+\infty\},\,a\leq b$ \\
&|& {\it expr} $\diamond$ {\it expr} && 
$\diamond\in\{\,+,-,\times,\slash\,\}$ \\
\\
{\it inst} &::=& 
   $X$ $\leftarrow$ {\it expr} && $X\in\V$ \\
&|& ${\it expr}\;\bowtie\;0\;?$ && 
$\bowtie\;\in\;\{\,=,\neq,<,\leq,\geq,>\,\}$\\
\end{tabular}
\caption{Syntax of our simple language.}
\label{syntax}
\end{myfig}

For the sake of presentation, we will only consider in this article a 
very simplified programming language focusing on manipulating
numerical variables.
We suppose that a program manipulates only a fixed, finite 
set of $n$ variables, $\V\deq\{V_1,\ldots,V_n\}$, 
with values within a perfect mathematical set, $\I\in\{\Z,\Q,\R\}$.
A program $P\in\P(\L\times{\it inst}\times\L)$ 
is a single control-flow graph where nodes are program points, in $\L$,
and arcs are labelled by instructions in {\it inst}.
We denote by $e$ the entry program point.
As described in Fig.~\ref{syntax}, 
only two types of instructions are allowed: assignments
$(X\leftarrow \text{{\it expr}})$ and tests $({\it expr}\;\bowtie\;0\;?)$,
where {\it expr\/} are numerical expressions and
$\bowtie$ is a comparison operator.
In the syntax of expressions, classical numerical constants have been replaced 
with {\em intervals\/} 
$[a,b]$ with constant bounds---possibly $+\infty$ or $-\infty$.
Such intervals correspond to a non-deterministic choice of a new value 
within the bounds each time the expression is evaluated. 
This will be key in defining the concept of 
{\em expression abstraction\/} in Sects.~3--5.
Moreover, interval constants appear naturally in programs that fetch input
values from an external environment, or when modeling rounding errors in
floating-point computations.

Affine forms play an important role in program analysis as they are easy
to manipulate and appear frequently as program invariants.
We enhance affine forms with the non-determinism of intervals by defining
{\em interval affine forms\/} as the expressions of the form:
$[a_0,b_0]+\sum_k\ ([a_k,b_k]\times V_k)$.

\subsection{Concrete Semantics of the Language}

\begin{myfig}
$\begin{array}{llll}
\lb{X}(\rho) &\deq& \{\;\rho(X)\;\} \\
\lb{[a,b]}(\rho) &\deq& \{\;x\in\I\;|\;a\leq x\leq b\;\}\\
\lb{e_1\diamond e_2}(\rho) &\deq&
\{\;x\diamond y\;|\;x\in\lb{e_1}(\rho),\,y\in\lb{e_2}(\rho)\;\}
\qquad  \diamond\in\{+,-,\times\}
\\
\lb{e_1/ e_2}(\rho) &\deq&
\{\;\trunc(x\slash y)\;|\;x\in\lb{e_1}(\rho),\,y\in\lb{e_2}(\rho),\,
y\neq 0\;\}
&\text{if\/ $\I=\Z$}
\\
\lb{e_1/ e_2}(\rho) &\deq&
\{\;x\slash y\;|\;x\in\lb{e_1}(\rho),\,y\in\lb{e_2}(\rho),\,
y\neq 0\;\}
&\text{if\/ $\I\neq\Z$}
\\\\
\lc{X\leftarrow e}(R) & \deq &
\{\;\rho[X\mapsto v]\;|\;\rho\in R,\;v\in\lb{e}(\rho)\;\}
\\
\lc{e\bowtie 0\;?}(R) & \deq &
\{\;\rho\;|\;
\rho\in R\text{ and }\exists\,v\in\lb{e}(\rho),\;v\bowtie 0\text{ holds}\;\}
\end{array}$
\caption{Concrete semantics.}
\label{concrete}
\end{myfig}

The {\em concrete semantics\/} of a program is the most precise
mathematical expression of its behavior.
Let us first define an {\em environment\/} as a function, in
$\V\rightarrow\I$, associating a value to each variable.
We choose a simple {\em invariant semantics\/}
that associates to each program point $l\in\L$ the set of all environments 
$\X_l\in\P(\V\rightarrow\I)$ that can hold when $l$ is reached.
Given an environment $\rho\in(\V\rightarrow\I)$, the semantics
$\lb{\expr}(\rho)$ of an expression $\expr$, shown in Fig.~\ref{concrete},
is the set of values
the expression can evaluate to.
It outputs a set to account for non-determinism.
When $\I=\Z$, the $\trunc$ function rounds the possibly non-integer 
result of the division towards an integer by {\em truncation}, as it is
common in most computer languages.
Divisions by zero are undefined, that is, return no result; 
for the sake of simplicity, we have not introduced any error state. 
The semantics of assignments and tests is defined by {\em transfer functions\/} 
$\lc{\mi{inst}}: \P(\V\rightarrow\I)\rightarrow\P(\V\rightarrow\I)$
in Fig.~\ref{concrete}.
The assignment transfer function returns environments
where one variable has changed its value ($\rho[V\mapsto x]$ denotes the 
function equal to $\rho$ on $\V\setminus\{V\}$ and that maps $V$ to $x$).
The test transfer function filters environments to keep only those that
{\em may\/} satisfy the test.
We can now define the semantics $(\X_l)_{l\in\L}$
of a program $P$ as the smallest solution of the following equation system:

\hfil$\left\{\begin{array}{llll}
\X_e & = & \quad V\rightarrow\I\\
\X_l & = & \displaystyle \bigcup_{(l',i,l)\in P}\; \lc{i}(\X_{l'})
\quad & \text{when $l\neq e$}
\end{array}\right.
\qquad (1)
$\hfil

\noindent
It describes the {\em strongest invariant\/} at each program point.

\subsection{Abstract Interpretation and Numerical Abstract Domains}

The concrete semantics is very precise but cannot be computed fully 
automatically by a computer.
We will only try to compute a sound overapproximation, that is,
a {\em superset\/} of the environments reached by the program.
We use Abstract Interpretation \cite{ai,ai3} to design such an approximation.

\paragraph{Numerical Abstract Domains.}
An analysis is parameterized by a numerical abstract domain that
allows representing and manipulating selected subsets of environments.
Formally it is defined as:
\begin{itemize}
\item\unskip a set of computer-representable {\em abstract\/} elements
$\D^\s$,
\item a {\em partial order\/} $\sqsubseteq^\s$ on $\D^\s$ to model the relative 
precision of abstract elements,
\item a monotonic {\em concretization\/} 
$\gamma: \D^\s\rightarrow \P(\V\rightarrow\I)$, that assigns a concrete property
to each abstract element,
\item a greatest element $\top^\s$ for $\sqsubseteq^\s$ 
such that $\gamma(\top^\s)=(\V\rightarrow\I)$,
\item {\em sound\/} and computable abstract versions 
$\lc{\mi{inst}}^\s$ of all transfer functions,
\item {\em sound\/} and computable abstractions $\cup^\s$ and $\cap^\s$ of 
$\cup$ and $\cap$,
\item a widening operator $\widen^\s$ if $\D^\s$ has infinite increasing chains.
\end{itemize}\unskip
The {\em soundness condition\/} for the abstraction 
$F^\s:(\D^\s)^n\rightarrow\D^\s$ of a $n-$ary operator
$F$ is: $F(\gamma(X^\s_1),\ldots,\gamma(X^\s_n))\subseteq
\gamma(F^\s(X^\s_1,\ldots,X^\s_n))$.
It ensures that $F^\s$ does not forget any of $F$'s behaviors.
It can, however, introduce spurious ones.

\paragraph{Abstract Analysis.}
Given an abstract domain, an abstract version $(1^\s)$ of the equation system
$(1)$ can be derived as:

\hfil$\left\{\begin{array}{llll}
\X^\s_e & = & \quad\top^\s\\
\X^\s_l & \sqsubseteq^\s & \displaystyle 
\sideset{}{^\s}\bigcup_{(l',i,l)\in P}\; \lc{i}^\s(\X^\s_{l'})
\quad & \text{when $l\neq e$}
\end{array}\right.
\qquad (1^\s)
$\hfil

\noindent
The soundness condition ensures that any solution of $(1^\s)$
satisfies $\forall\,l\in\L,\;\gamma(\X^\s_l)\supseteq \X_l$.
The system can be solved by iterations, using
a widening operator $\widen^\s$ to ensure termination.
We refer the reader to Bourdoncle \cite{widen} for an in-depth description
of possible iteration strategies.
The computed $\X^\s_l$ is almost never the best 
abstraction---if it exists---of the concrete solution $\X_l$.
Unavoidable losses of precision come from the use of convergence acceleration
$\widen^\s$,
non-necessarily best abstract transfer functions, and the fact that the 
composition of best abstractions is generally not a best abstraction.
This last issue explains why even the simplest semantics-preserving 
program transformations can drastically affect the quality of a static 
analysis.

\paragraph{Existing Numerical Domains.}
There exists many numerical abstract domains.
We will be mostly interested in those able to express variable bounds.
Such abstract domains include the well-known interval domain \cite{interv} 
(able to express invariants of the form $\bigwedge_i\;V_i\in[a_i,b_i]$),
and the polyhedron domain \cite{poly} (able to express affine inequalities
$\bigwedge_i\;\sum_{j} \alpha_{ij} V_i \geq \beta_j$).
More recent domains, in-between these two in terms of cost and precision,
include the octagon domain \cite{mine:oct}
($\bigwedge_{ij}\;\pm V_i\pm V_j\leq c_{ij}$),
the octahedron domain \cite{octahedra}
($\bigwedge_j\;\sum_{i} \alpha_{ij} V_i \geq \beta_j$
where $\alpha_{ij}\in\{-1,0,1\}$),
and the Two Variable Per Inequality domain \cite{tvpli}
($\bigwedge_i\;\alpha_i V_{k_i}+\beta_i V_{l_i}\leq c_i$).

\section{Incorporating Symbolic Methods}
\label{symbolicsect}

\vspace*{-0.2cm}

We suppose that we are given a numerical abstract domain $\D^\s$.
The gist of our method is to replace, in the abstract transfer functions
$\lc{X\leftarrow e}^\s$ and $\lc{e\bowtie 0\;?}^\s$,
each expression $e$ with another one $e'$, in a sound way.

\paragraph{Partial Order on Expressions.}
To define formally the notion of sound expression abstraction,
we first introduce an approximation order $\expleq$ on expressions.
A natural choice is to consider the point-wise ordering of the 
concrete semantics $\lb{\cdot}$ defined in Fig.~\ref{concrete}, that is:
$e_1\expleq e_2\diff\forall\,\rho\in(\V\rightarrow\I),\;\lb{e_1}(\rho)
\subseteq\lb{e_2}(\rho)$.
However, requiring the inclusion to hold for {\em all\/} environments
is quite restrictive.
More aggressive expression transformations can be enabled by only requiring 
soundness with respect to selected sets of environments.
Our partial order $\expleq$ is now defined ``up to'' a
set of environments $R\in\P(\V\rightarrow\I)$:
\begin{defn}
$R\models e_1\expleq e_2\diff\forall\,\rho\in R,\;\lb{e_1}(\rho)
\subseteq\lb{e_2}(\rho).$
\end{defn}
We denote by $R\models e_1=e_2$ the associated equality relation.

\paragraph{Sound Symbolic Transformations.}
We wish now to abstract some transfer function, 
{\em e.g.}, $\lc{V\leftarrow e}$, on an abstract environment
$R^\s\in\D^\s$.
The following theorem states that, if $e'$ overapproximates $e$ on 
$\gamma(R^\s)$, it is sound to replace $e$ with $e'$ in the abstract
transfer functions:
\begin{thm}
\label{mainthm}
If\/ $\gamma(R^\s)\models e\expleq e'$, then:
\begin{itemize}
\item[$\bullet$]\unskip
$(\lc{V\leftarrow e}\circ\gamma)(R^\s)
 \subseteq (\gamma\circ\lc{V\leftarrow e'}^\s)(R^\s)$,
\item[$\bullet$]
$(\lc{e\bowtie 0\;?}\circ\gamma)(R^\s)
 \subseteq (\gamma\circ\lc{e'\bowtie 0\;?}^\s)(R^\s)$.
\end{itemize}
\end{thm}


\section{Expression Linearization}
\label{linearizationsect}

\vspace*{-0.2cm}

Our first symbolic transformation is an abstraction of arbitrary expressions
into interval affine forms $i_0+\sum_k(i_k\times V_k)$, where the
$i$'s stand for intervals.

\subsection{Definitions}

\paragraph{Interval Affine Form Operators.}
We first introduce a few operators to manipulate interval affine forms in
a symbolic way.
Using the classical interval arithmetic operators---denoted with a $\Int$
superscript---we can define point-wisely the addition $\boxplus$ and 
subtraction $\boxminus$ of affine forms, as well as the multiplication 
$\boxtimes$ and division $\boxslash$ of an affine form by a constant interval:
\begin{defn}~
\begin{itemize}
\item[$\bullet$]\unskip
$(i_0+\sum_k i_k\times V_k)\;\boxplus\;(i_0'+\sum_k i'_k\times V_k)
\deq (i_0+^\Int i'_0)+\sum_k (i_k +^\Int i'_k)\times V_k$,
\item[$\bullet$]
$(i_0+\sum_k i_k\times V_k)\;\boxminus\;(i_0'+\sum_k i'_k\times V_k)
\deq (i_0-^\Int i'_0)+\sum_k (i_k -^\Int i'_k)\times V_k$,
\item[$\bullet$]
$i\;\boxtimes\;(i_0+\sum_k i_k\times V_k)
\deq (i\times^\Int i_0)+\sum_k\,(i\times^\Int i_k)\times V_k$,
\item[$\bullet$]
$(i_0+\sum_k i_k\times V_k)\;\boxslash\; i
\deq (i_0/^\Int\, i)+\sum_k\,(i_k/^\Int\, i)\times V_k$.
\end{itemize}
where the interval arithmetic operators are defined classically as:
\begin{itemize}
\item[$\bullet$]\unskip $[a,b] +^\Int [a',b'] \deq [a+a',b+b']$,
\qquad $\bullet$ $[a,b] -^\Int [a',b'] \deq [a-b',b-a']$,
\item[$\bullet$] $[a,b] \times^\Int [a',b'] \deq 
[\min (a a',a b',b a',b b'),
 \max (a a',a b',b a',b b')]$,
\item[$\bullet$] $[a,b] /^\Int [a',b'] \deq \\
\left\{\begin{array}{ll}
{}[-\infty,+\infty] &
\text{ if\/ $0\in[a',b']$}
\\
{}[\min (a/a',a/b',b/a',b/b'),\;
   \max (a/a',a/b',b/a',b/b')] &
\text{ when\/ $\I\neq\Z$}
\\
{}[\lfloor \min (a/a',a/b',b/a',b/b')\rfloor,\;
   \lceil  \max (a/a',a/b',b/a',b/b')\rceil] &
\text{ when\/ $\I=\Z$}
\end{array}\right.$
\end{itemize}
\end{defn}
The following theorem states that these operators are always sound and,
in some cases, complete---{\em i.e.}, $\expleq$ can be replaced by $=$:
\begin{thm}
\label{sound1thm}
For all interval affine forms $l_1$, $l_2$ and interval $i$, we have:

\begin{tabular}{llll}
$\bullet\;$ & $\I^\V\models l_1+l_2 = l_1\boxplus l_2$, &
$\bullet\;$ & $\I^\V\models l_1-l_2 = l_1\boxminus l_2$, \\
$\bullet$ & $\I^\V\models i\times l_1 = i\boxtimes l_1$, if\/ $\I\neq\Z$,&
$\bullet$ & $\I^\V\models i\times l_1 \expleq i\boxtimes l_1$, otherwise,\\
$\bullet$ & $\I^\V\models l_1 / i = l_1\boxslash i$, if\/ $\I\neq\Z$ and 
$0\notin i$, \qquad &
$\bullet$ & $\I^\V\models l_1 / i \expleq l_1\boxslash i$, otherwise.\\
\end{tabular}
\end{thm}
When $\I=\Z$, we must conservatively round upper and lower bounds respectively
towards $+\infty$ and $-\infty$ to ensure that Thm.~\ref{sound1thm} holds.
The non-exactness of the multiplication and division can then lead
to some precision degradation.
For instance, $(X\boxslash 2)\boxtimes 2$ evaluates to $[0,2]\times X$
as, when computing $X\boxslash 2$, the non-integral value 
$1/2$ must be abstracted into the integral interval $[0,1]$.
One solution is to perform all computations in $\R$, keeping in mind that, due
to truncation, $l/[a,b]$ should be interpreted when $0\notin[a,b]$ as
$(l\,\boxslash\,[a,b])\;\boxplus\;[-1+x,1-x]$, where $x=1/\min(|a|,|b|)$.
We then obtain the more precise result $X+[-1,1]$.

We now  introduce a so-called ``intervalization'' operator, $\iota$, to 
abstracts interval affine forms into intervals.
Given an abstract environment, it evaluates the affine form using interval 
arithmetics.
Suppose that $\D^\s$ provides us with projection operators 
$\pi_k:\D^\s\rightarrow \P(\I)$ able to return
an interval overapproximation for each variable $V_k$.
We define $\iota$ as:
\begin{defn}
$\iota(i_0+\sum_k (i_k\times V_k))(R^\s)\deq
i_0\;+^\Int\;\sum_k^\Int\;(i_k\times^\Int \pi_k(R^\s))$,\\
where each $\pi_k(R^\s)$ is an interval containing
$\{\;\rho(V_k)\;|\;\rho\in\gamma(R^\s)\;\}$.
\end{defn}
The following theorem states that $\iota$ is a sound operator 
with respect to $R^\s$:
\begin{thm} 
\label{sound2thm}
$\gamma(R^\s)\models l\expleq\iota(l)(R^\s)$.
\end{thm}
As $\pi_k$ performs a non-relational abstraction, $\iota$
incurs a loss of precision whenever $\D^\s$ is a relational domain.
Consider, for instance $R^\s$ such that 
$\gamma(R^\s)=\{\;\rho\in(\{V_1,V_2\}\rightarrow[0,1])\;|\;
\rho(V_1)=\rho(V_2)\;\}$.
Then, $\lb{\iota(V_1-V_2)(R^\s)}$ is the constant function $[-1,1]$ while
$\lb{V_1-V_2}$ is $0$.

\paragraph{Linearization.}
The linearization $\lp{e}(R^\s)$ of an arbitrary expression $e$
in an abstract environment $R^\s$ can now be defined by structural
induction as follows:
\begin{defn}
\label{lineardef}~
\begin{itemize}
\item[$\bullet$]\unskip $\lp{V}(R^\s)\deq [1,1]\times V$,
\hfil $\bullet$ $\lp{[a,b]}(R^\s)\deq [a,b]$,
\item[$\bullet$] $\lp{e_1+e_2}(R^\s)\deq \lp{e_1}(R^\s)\;\boxplus\;\lp{e_2}(R^\s)$,
\item[$\bullet$] $\lp{e_1-e_2}(R^\s)\deq \lp{e_1}(R^\s)\;\boxminus\;\lp{e_2}(R^\s)$,
\item[$\bullet$] $\lp{e_1/e_2}(R^\s)\deq \lp{e_1}(R^\s)\;\boxslash\;
\iota(\lp{e_2}(R^\s))(R^\s)$,
\item[$\bullet$] $\lp{e_1\times e_2}(R^\s)\deq
\left\{\begin{array}{ll}
\text{either } & \iota(\lp{e_1}(R^\s))(R^\s)\;\boxtimes\;\lp{e_2}(R^\s)
\\
\text{or }     & \iota(\lp{e_2}(R^\s))(R^\s)\;\boxtimes\;\lp{e_1}(R^\s)
\end{array}\right.$
\; (see Sect.~\ref{multsect})
\end{itemize}
\end{defn}
The $\iota$ operator is used to deal with non-linear constructions:
the right argument of a division and either argument of a multiplication 
are intervalized.
As a consequence of Thms.~\ref{sound1thm} and \ref{sound2thm}, 
our linearization is sound:
\begin{thm}
\label{sound3thm}
$\gamma(R^\s)\models e \expleq\lp{e}(R^\s).$
\end{thm}
Obviously, $\lp{\cdot}$ generally incurs a loss of precision with respect to
$\expleq$.
Also, $\lp{e}$ is not monotonic in its $e$ argument.
Consider for instance $X/X$ in the environment $R^\s$ such that
$\pi_X(R^\s)=[1,+\infty]$.
Although $\gamma(R^\s)\models X/X\expleq[1,1]$,
we do not have
$\gamma(R^\s)\models \lp{X/X}(R^\s)\expleq \lp{[1,1]}(R^\s)$
as $\lp{X/X}(R^\s)=[0,1]\times X$.
It is important to note that there is no
useful notion of {\em best abstraction\/} of expressions for $\expleq$.

\subsection{Integration With a Numerical Abstract Domain}

Given an abstract domain, $\D^\s$, we can now derive a new abstract domain
with linearization, $\D^\s_\L$, identical to  $\D^\s$ except for the following
transfer functions:

\hfil$
\begin{array}{lll}
\lc{V\leftarrow e}^\s_\L(R^\s) &\deq& \lc{V\leftarrow \lp{e}(R^\s)}^\s(R^\s)\\
\lc{e\bowtie 0\;?}^\s_\L(R^\s) &\deq& \lc{\lp{e}(R^\s)\bowtie 0\;?}^\s(R^\s)\\
\end{array}
$\hfil

\noindent
The soundness of these transfer functions is guaranteed by
Thms.~\ref{mainthm} and \ref{sound3thm}.

\paragraph{Application to the Interval Domain.}
As all non-relational domains, the interval domain \cite{interv}, is
not able to exploit the fact that the same variable occurs several times in 
an expression.
Our linearization performs some symbolic simplification, and so, is able to
partly correct this problem.
Consider, for instance, the assignment $\lc{Y\leftarrow 3\times X-X}$
in an abstract environment such that $X\in[a,b]$.
The regular interval domain $\D^\Int$ will assign $[3a-b,3b-a]$ to $Y$, 
while $\D^\Int_\L$ will assign $[2a,2b]$ as $\lp{3\times X-X}(R^\s)=2\times X$.
This last answer is strictly more precise whenever $a\neq b$.
Using the exactness of Thm.~\ref{sound1thm}, one can prove that, when 
$\I\neq\Z$, the assignment in $\D^\Int_\L$ is always more precise than in
$\D^\Int$.
This may not be the case for a test, or when $\I=\Z$.

\paragraph{Application to the Octagon Domain.}
The octagon domain \cite{mine:oct}
is more precise than the interval one, but it is more complex.
As a consequence, it is quite difficult to design abstract transfer functions
for non-linear expressions.
This problem can be solved by using our linearization in combination with
the efficient and rather precise interval affine form abstract transfer 
functions proposed in our previous work \cite{mine:these}.
The octagon domain with linearization is able to prove, for instance,
that, after the assignment $X\leftarrow T\times Y$ in an environment such
that $T\in[-1,1]$, we have $-Y\leq X\leq Y$.

\paragraph{Application to the Polyhedron Domain.}
The polyhedron domain \cite{poly} 
is more precise than the octagon domain but cannot deal with full interval 
affine forms---only the constant coefficient may safely be an interval.
To solve this problem, we introduce a function $\mu$ to abstract interval
affine forms further by making all variable coefficients singletons.
For the sake of conciseness, we give a formula valid only for 
$\I\neq Z$ and finite interval bounds:
\begin{defn}~\\
$\begin{array}[t]{l}
\mu\left([a_0,b_0]+\sum_k [a_k,b_k]\times V_k\right)(R^\s) \deq \\
\left([a_0,b_0] +^\Int 
\sum_k^\Int\,[(a_k-b_k)/2,(b_k-a_k)/2]\times^\Int \pi_k(R^\s)\right)
\;+\;\sum_k\,((a_k+b_k)/2)\times V_k
\end{array}$
\end{defn}
$\mu$ works by ``distributing'' the weight $b_k-a_k$ of
each variable coefficient into the constant component, using variable bounds
information from $R^\s$.
One can prove that $\mu$ is sound, that is, 
$\gamma(R^\s)\models l\expleq\mu(l)R^\s$.

\paragraph{Application to Floating-Point Arithmetics.}
Real-life programming languages do not manipulate rationals or reals, but
floating-point numbers, which are much more difficult to abstract.
Pervasive rounding must be taken into account.
As most classical properties of arithmetic operators are no longer true, it is
generally not safe to feed floating-point expressions to relational domains.
One solution is to convert such expressions into real-valued expressions by
making rounding explicit.
Rounding is highly non-linear but can be abstracted using intervals.
For instance, $X+Y$ in the floating-point world can be abstracted into
$[1-\epsilon_1,1+\epsilon_1]\times X+[1-\epsilon_1,1+\epsilon_1]\times Y+
[-\epsilon_2,\epsilon_2]$ using small constants $\epsilon_1$ and
$\epsilon_2$ modeling, respectively, relative and absolute errors.
This fits in our linearization framework which can be extended to treat soundly
floating-point arithmetics.
We refer the reader to related work \cite{mine:float} for more information.

\subsection{Multiplication Strategies}
\label{multsect}

When encountering a multiplication $e_1\times e_2$ and neither 
$\lp{e_1}(R^\s)$ nor $\lp{e_2}(R^\s)$ evaluates to an interval, we must
intervalize either argument.
Both choices are valid, but influence greatly the precision of the result.

\paragraph{All-Cases Strategy.}
A first idea is to try both choices for each multiplication; we get
a {\em set\/} of linearized expressions.
We have no notion of greatest lower bound on expressions, so, we must
evaluate a transfer function for all expressions in parallel, and take
the intersection $\cap^\s$ of the resulting abstract elements in $\D^\s$.
Unfortunately, the cost is exponential in the number of multiplications
in the original expression,
hence the need for deterministic strategies that always select {\em one\/} 
interval affine form.

\paragraph{Interval-Size Strategy.}
A simple strategy is to intervalize the affine form that will
yield the narrower interval.
This greedy approach tries to limit the amplitude of the
non-determinism introduced by multiplications.
The extreme case holds when the amplitude of one interval is zero, meaning that
the sub-expression is semantically a constant; intervalizing it will not
result in any precision loss.
Finally, note that the {\em relative\/} amplitude $(b-a)/|a+b|$ may be more
significant than the absolute amplitude $b-a$ if we want to intervalize
preferably expressions that are constant up to some small relative rounding
error.

\paragraph{Simplification-Driven Strategy.}
Another idea is to maximize the amount of simplification by
not intervalizing, when possible, sub-expressions containing variables
appearing in other sub-expressions.
For instance, in $X-(Y\times X)$, $Y$ will be intervalized to
yield $[1-\max Y,1-\min Y]\times X$.
Unlike the preceding greedy approach, this strategy is global and treats the 
expression as a whole.

\paragraph{Homogeneity Strategy.}
We now consider the linear interpolation of Fig.~\ref{interpol}.
In order to achieve the best precision, it is important to intervalize
$X$ in both multiplications.
This yields $T\leftarrow [0,1]\times Y+[0,1]\times Z$ and we are able to
prove that $T\geq 0$---however, we find that $T\leq 0.3$ while in fact
$T\leq 0.2$.
The interval-size strategy would choose to intervalize $Y$ and $Z$ that have 
smaller range than $X$, which yields the imprecise
assignment $T\leftarrow [-0.2,0.1]\times X+[0,0.2]$.
Likewise, the simplification-driven strategy may choose to keep $X$
that appears in two sub-expressions and also intervalize
both $Y$ and $Z$.
To solve this problem, we propose to intervalize 
the smallest set of variables that makes the expression homogeneous, that is,
arguments of $+$ and $-$ operators should have the same degree.
In order to make the $(1-X)$ sub-expression homogeneous, $X$ is intervalized.
This last strategy is quite robust: it keeps working if we change the 
assignment into the equivalent $T\leftarrow X\times Y-X\times Z+Z$, or
if we consider bi-linear interpolations or interpolations with
normalization coefficients.

\subsection{Concluding Remark}
Our linearization is not equivalent to a static program transformation.
To cope with non-linearity as best as we can, we exploit the information 
dynamically inferred by the analysis:
first, in the intervalization $\iota$, then, in the multiplication strategy.
Both algorithms take as argument the current numerical abstract environment 
$R^\s$.
As, dually, the linearization improves the precision of the next 
computed abstract element, the dynamic nature of our approach ensures
a positive feed-back.


\section{Symbolic Constant Propagation}
\label{constantsect}

\vspace*{-0.2cm}

The automatic symbolic simplification implied by our linearization 
allows us to gain much precision when dealing with complex expressions, 
without the burden of designing new abstract domains tailored for them.
However, the analysis is still sensitive to program transformations that
decompose expressions and introduce new temporary variables---such as
common sub-expression elimination or register spilling.
In order to be immune to this problem, one must generally use an expressive,
and so, costly, {\em relational\/} domain.
We propose an alternate, lightweight solution based on a kind of
constant domain that tracks assignments dynamically and propagate symbolic
expressions within transfer functions.

\subsection{The Symbolic Constant Domain}

\paragraph{Enriched Expressions.}
We denote by $\S$ the set of all syntactic expressions, enriched with
one element $\top^\S$ denoting `any value.'
The flat ordering $\sqsubseteq^\S$ is defined as
$X\sqsubseteq^\S Y\iff Y=\top^\S\text{ or }X=Y$.
The concrete semantics $\lb{\cdot}$ of Fig.~\ref{concrete} is extended
to $\S$ as $\lb{\top^\S}(\rho)=\I$.
We also use two functions on expression trees:
$\occ:\S\rightarrow \P(\V)$ that returns the set of variables occurring
in an expressing, and $\subst:\S\times\V\times\S\rightarrow\S$
that substitutes, in its first argument, every occurrence of a given variable 
by its last argument.
Their definition on non$-\top^\S$ elements is quite standard and we do not
present it here.
They are extended to $\S$ as follows: $\occ(\top^\S)\deq\emptyset$,
$\subst(e,V,\top^\S)$ equals $e$ when $V\notin\occ(e)$ and $\top^\S$
when $V\in\occ(e)$.
\looseness=-1

\paragraph{Abstract Symbolic Environments.}
The {\em symbolic constant domain\/} is the set
$\D^\S\deq\V\rightarrow \S$ restricted as follows:
there must be no cyclic dependencies in a map $S^\S\in\D^\S$, that is, 
pair-wise distinct variables $V_1,\ldots,V_n$ such that 
$\forall i,\,V_i\in\occ(S^\S(V_{i+1}))$ and $V_n\in\occ(S^\S(V_1))$.
The partial order $\sqsubseteq^\S$ on $\D^\S$ is the point-wise extension of
that on $\S$.
Each element $S^\S\in\D^\S$ represents the set of environments compatible with
the symbolic information:
\begin{defn}
$\gamma^\S(S^\S)\deq 
\{\;\rho\in(\V\rightarrow\I)\;|\;
\forall k,\;\rho(V_k)\in\lb{S^\S(V_k)}(\rho)\;\}.$
\end{defn}

\paragraph{Main Theorem.} 
Our approach relies on the fact that
applying a substitution from $S^\S$ to any expression
is sound with respect to $\gamma^\S(S^\S)$:
\begin{thm}
$\forall e,V,S^\S,\;\gamma^\S(S^\S)\models e \expleq \subst(e,V,S^\S(V))$.
\end{thm}

\paragraph{Abstract Operators.}
We now define the following operators on $\D^\S$:

\vspace*{-0.2cm}
\begin{defn}~
\begin{itemize}
\item[$\bullet$]\unskip 
$\lc{V\leftarrow e}^\S(S^\S)(V_k)\deq
\left\{\begin{array}{ll}
\subst(e,\,V,\,S^\S(V)) & \text{if\/ $V=V_k$}\\
\subst(S^\S(V_k),\,V,\,S^\S(V)) \quad & \text{if\/ $V\neq V_k$}
\end{array}\right.$
\item[$\bullet$] $\lc{e\bowtie 0\;?}^\S(S^\S)\deq  S^\S$,
\item[$\bullet$] $(S^\S\cup^\S T^\S)(V_k)\deq
\left\{\begin{array}{ll}
S^\S(V_k) \quad & \text{ if\/ $S^\S(V_k)=T^\S(V_k)$}\\
\top^\S &\text{ otherwise}
\end{array}\right.$
\item[$\bullet$] $S^\S\cap T^\S \deq S^\S$
\end{itemize}
\end{defn}
\vspace*{-0.2cm}
Our assignment $V\leftarrow e$ first substitutes $V$ with $S^\S(V)$
in $S^\S$ and $e$ before adding the information that $V$ maps to the 
substituted $e$.
This is necessary to remove all prior information on $V$ (no
longer valid after the assignment) and prevent the apparition of dependency 
cycles.
As we are only interested in propagating assignments, tests are abstracted
as the identity, which is sound but coarse.
Our union abstraction only keeps syntactically equal expressions.
This corresponds to the least upper bound with respect to $\sqsubseteq^\S$. 
Our intersection keeps only the information of the left argument.
All these operators respect the non-cyclicity condition.
Note that one could be tempted to refine the intersection by mixing 
information from the left and right arguments in order to minimize the
number of variables mapping to $\top^\S$.
Unfortunately, careless mixing may break the non-cyclicity condition. 
We settled, as a simpler but safe solution, to keeping the left argument.
Finally, we do not need any widening: at each abstract iteration, unstable 
symbolic expressions are directly replaced with $\top^\S$ when applying
$\cup^\S$, and so, become stable.

\subsection{Integration With a Numerical Abstract Domain}

Given a numerical abstract domain $\D^\s$, the domain $\D^{\s\times\S}$ is
obtained by combining $\D^\s_\L$ with $\D^\S$ the following way:
\begin{defn}~
\begin{itemize}
\item[$\bullet$]\unskip $\D^{\s\times\S}\deq\D^\s\times\D^\S$,
\item[$\bullet$] $\sqsubseteq^{\s\times\S}$, $\cup^{\s\times\S}$ and
$\cap^{\s\times\S}$ are defined pair-wise, and
$\widen^{\s\times\S}\deq\widen^\s\times\cup^\S$,
\item[$\bullet$] $\gamma^{\s\times\S}(R^\s,S^\S)\deq 
\gamma^\s(R^\s)\cap\gamma^\S(S^\S)$,
\item[$\bullet$] $\lc{V\leftarrow e}^{\s\times\S}(R^\s,S^\S)\deq
(\lc{V\leftarrow \strat(e,S^\S)}^\s_\L(R^\s),\,\lc{V\leftarrow e}^\S(S^\S))$
\item[$\bullet$] $\lc{e\bowtie 0\;?}^{\s\times\S}(R^\s,S^\S)\deq
(\lc{\strat(e,S^\S)\bowtie 0\;?}^\s_\L(R^\s),\,\lc{e\bowtie 0\;?}^\S(S^\S))$
\end{itemize}
Where $\strat(e,S^\S)$ is a\/ {\em substitution strategy} that
may perform sequences of substitutions of the form 
$f\mapsto\subst(f,V,S^\S(V))$ in $e$, for any variables $V$.
\end{defn}
All information in $\D^\S$ and $\D^\s$ are computed independently, except
that the symbolic information is used in the transfer functions
for $\D^\s_\L$.
The next section discusses the choice of a strategy $\strat$.
Note that, although we chose in this presentation to abstract the semantics of
Fig.~\ref{concrete}, our construction can be used on any class of expressions,
including floating-point and non-numerical expressions.

\subsection{Substitution Strategies}
\label{subststrat}

Any sequence of substitutions extracted from the current symbolic constant
information is sound, but some give better results than others.
As for the intervalization of Sect.~\ref{multsect}, 
we rely on carefully designed strategies.

\paragraph{Full Propagation.}
Thanks to the non-cyclicity of elements $S^\S\in\D^\S$, we can
safely perform all substitutions $f\mapsto\subst(f,V,S^\S(V))$
for all $V$ in any order, and reach a normal form.
This gives a first basic substitution strategy.
However, because our goal is to perform linearization-driven simplifications,
it is important to avoid substituting with variable-free expressions
or we may lose correlations between multiple occurrences of variables.
For instance, full substitution in the assignment $Z\leftarrow X-0.5\times Y$ 
with the environment $S^\S=[X\mapsto [0,1],\;Y\mapsto X]$ results in 
$Z\leftarrow [0,1]-0.5\times[0,1]$, and so, $Z\in[-0.5,1]$.
Avoiding variable-free substitutions, this gives $Z\leftarrow X-0.5\times X$, 
and so, $Z\in[0,0.5]$, which is more precise.
This refined strategy also succeeds in proving that $Y\in[0,20]$ in the 
example of Fig.~\ref{absolute} by substituting $Y$ with $X$ in the test 
$Y\leq 0$.

\paragraph{Enforcing Determinism and Linearity.}
Non-determinism in expressions is a major source of precision loss.
Thus, a strategy is to avoid substituting $V$ with $S^\S(V)$ whenever 
$\#(\lb{S^\S(V)}\circ\gamma)(X^\s)>1$.
As this property is not easily computed, we propose the following sufficient
syntactic criterion: $S^\S(V)$ should not be $\top^\S$ nor contain a 
non-singleton interval.
This strategy gives the expected result in the example of Fig.~\ref{absolute}.
Likewise, one may wish to avoid substituting with non-linear expressions, as
they must be subsequently intervalized, which is a cause of precision loss.
However, disabling too many substitutions may prevent the 
linearization step to exploit correlations.
Suppose that we break the last assignment of Fig.~\ref{interpol}
in three parts: 
$U\leftarrow X\times Y;$ $V\leftarrow (1-X)\times Z;$ $T\leftarrow U-V$.
Then, the interval domain with linearization and symbolic constant propagation 
will not be able to prove that $T\in[0,0.3]$ unless we allow
substituting, in $T$, $U$ and $V$ with their {\em non-linear\/} symbolic value.

\paragraph{Gaining More Precision.}
More precision can be achieved by slightly altering the definition of
$\D^{\s\times\S}$.
A simple but effective idea is to allow several strategies, compute several
transfer functions in $\D^\s$ in parallel, and take the abstract intersection 
$\cap^\s$ of the results.
Another idea is to perform reductions from $\D^\S$ to $\D^\s$ after each
transfer function: $X^\s$ is replaced with
$\lc{V_k-S^\S(V_k)=0\;?}^\s(X^\s)$ for some $k$.
Reductions can be iterated to increase the precision, following
Granger's local iterations scheme \cite{localiter}.


\section{Application to the \astree\ Analyzer}
\label{astreesect}

\vspace*{-0.2cm}

\astree\ is an efficient static analyzer focusing on the
detection of run-time errors for programs written in a subset of the 
C programming language, excluding recursion, dynamic memory allocation and
concurrent executions.
It aims towards a degree of precision sufficient to actually {\em prove\/} 
the absence of run-time errors.
This is achieved by specializing the analyzer towards specific program
families, introducing various abstract domains, and setting iteration
strategy parameters. 
Currently, the considered family of programs is that of 
safety, critical, embedded, fly-by-wire avionic software,
featuring large reactive loops running for billions of iterations,
thousands of global state variables, and pervasive floating-point 
arithmetics.
We refer the reader to \cite{magic2} for more detailed informations on \astree.

\paragraph{Integrating the Symbolic Methods.}
\astree\ uses a partially reduced product of several 
numerical abstract domains, together with both our two symbolic enhancement 
methods.
Relational domains, such as the octagon \cite{mine:oct} or digital
filtering \cite{filters} domains, rely on the linearization to abstract complex 
floating-point expressions into interval affine forms on reals.
The interval domain is refined by combining three versions of each transfer 
function. 
Firstly, using the expression unchanged. 
Secondly, using the linearized expression.
Thirdly, applying symbolic constant propagation followed by
linearization.
We use the simplification-driven multiplication strategy, as well as the
full propagation strategy---not propagating variable-free expressions.

\paragraph{Experimental  Results.}
We present analysis results on a several programs.
All the analyses have been carried on an 64-bit AMD Opteron 248 
(2 GHz) workstation running Linux, using a single processor.
The following table compares the precision and efficiency of \astree\ before 
and after enabling our two symbolic methods:

\begin{center}
\begin{tabular}{|r|r|c|r|c||r|c|r|c|}
\cline{2-9}
\multicolumn{1}{c}{} &
\multicolumn{4}{|c||}{{\bf without enhancements}} &
\multicolumn{4}{|c|}{{\bf with enhancements}}
\\
\hline
\multicolumn{1}{|c|}{\begin{tabular}{c}code size\\in lines\end{tabular}} &
\multicolumn{1}{c|}{\begin{tabular}{c}analysis\\time\end{tabular}} &
\multicolumn{1}{c|}{\begin{tabular}{c}nb. of\\iters.\end{tabular}} &
\multicolumn{1}{c|}{memory} &
\multicolumn{1}{c||}{alarms} & 
\multicolumn{1}{c|}{\begin{tabular}{c}analysis\\time\end{tabular}} &
\multicolumn{1}{c|}{\begin{tabular}{c}nb. of\\iters.\end{tabular}} &
\multicolumn{1}{c|}{memory} &
\multicolumn{1}{c|}{alarms}
\\
\hline\hline

    370 &   1.8s   & 17  &   16 MB &      0
        &   3.1s   & 17  &   16 MB &      0  \\
  9 500 &    90s   & 39  &   80 MB &      8 
        &   160s   & 39  &   81 MB &      8  \\
 70 000 & 2h 40mn  & 141 &  559 MB &    391
        & 1h 16mn  & 44  &  582 MB & {\bf 0} \\
226 000 & 11h 16mn & 150 &  1.3 GB &    141
        & 6h 36mn  & 86  &  1.3 GB &      1  \\
400 000 & 22h 9mn  & 172 &  2.2 GB &    282
        & 13h 52mn & 96  &  2.2 GB & {\bf 0} \\
\hline
\end{tabular}
\end{center}

The precision gain is quite impressive as up to hundreds 
of alarms are removed.
In two cases, this increase in precision is sufficient to achieve zero alarm,
which actually {\em proves\/} the absence of run-time errors.
Moreover, the increase in memory consumption is negligible.
Finally, in our largest examples, our enhancement methods {\em save\/} 
analysis time:
although each abstract iteration is more costly (up to
$25 \%$) this is compensated by the reduced number of iterations needed
to stabilize our invariants as a smaller state space is explored.

\paragraph{Discussion.}
It is possible to use the symbolic constant propagation also in
relational domains, but this was not needed in our examples to remove alarms.
Our experiments show that, even though the linearization and 
constant propagation techniques on intervals are not as robust as 
fully relational abstract domains, they are quite versatile thanks to 
their parametrization in terms of strategies, and much simpler to implement 
than even a simple relational abstract domain.
Moreover, our methods exhibit a near-linear time and memory cost, which is
much more efficient than relational domains.


\section{Conclusion}
\label{conclusionsect}

\vspace*{-0.2cm}

We have proposed, in this article, two techniques, called linearization and
symbolic constant propagation, that can be combined together to improve the
precision of numerical abstract domains.
In particular, we are able to compensate for the lack of non-linear transfer 
functions in the polyhedron and octagon domains, and for a weak or inexistent 
level of relationality in the octagon and interval domains.
Finally, they help making abstract domains robust against program
transformations.
Thanks to their parameterization in terms of strategies, they can be 
finely tuned to take into account semantics as well as syntactic program
features.
They are also very lightweight in terms of both analysis and development costs.
We found out that, in many cases, it is easier and faster to design a couple of
linearization and symbolic propagation strategies to solve a local loss
of precision in some program, while keeping the interval abstract domain,
than to develop a specific relational abstract domain able to represent the 
required local properties.
Strategies also proved reusable on programs belonging to the same family.
Practical results obtained within the \astree\ static analyzer show that
our methods both increase the precision and save analysis time.
They were key in proving the absence of run-time errors in
real-life critical embedded avionics software.

\paragraph{Future Work.}
Because the precision gain strongly depends upon the multiplication
strategy used in our linearization and the propagation strategy 
used in the symbolic constant domain, a natural extension of our work
is to try and design new such strategies, adapted to different
{\em practical\/} cases.
A more challenging task would be to provide {\em theoretical\/} guarantees that
some strategies make abstract domains immune to given classes of program
transformations.

\paragraph{Acknowledgments.}
We would like to thank all the former and present members of the
\astree\ team:
B.~Blanchet, P.~Cousot, R.~Cousot, J.~Feret, L.~Mauborgne, D.~Monniaux and 
X.~Rival.
We would also like to thank
the anonymous referees for their useful comments.


\bibliographystyle{plain}

\end{document}